\documentclass[aps,prx,twocolumn,nofootinbib,groupedaddress,superscriptaddress,longbibliography]{revtex4-2}

\usepackage{amssymb}
\usepackage{multirow}
\usepackage{graphicx}
\usepackage{amsmath}
\usepackage{color}
\usepackage{mathrsfs}
\usepackage{float}
\usepackage{indentfirst}
\usepackage{mathrsfs}
\usepackage{float}
\usepackage{indentfirst}
\usepackage{textcomp}
\usepackage{comment}
\usepackage{mathtools}
\usepackage{natbib,hyperref}
\usepackage{soul}

\begin{document}
\title{Non-stabilizerness Entanglement Entropy: a measure of hardness in the classical simulation of quantum many-body systems}

\author{Jiale Huang} \thanks{These authors contributed equally to this work.}
\affiliation{Key Laboratory of Artificial Structures and Quantum Control (Ministry of Education),  School of Physics and Astronomy, Shanghai Jiao Tong University, Shanghai 200240, China}

\author{Xiangjian Qian}  \thanks{These authors contributed equally to this work.}
\affiliation{Key Laboratory of Artificial Structures and Quantum Control (Ministry of Education),  School of Physics and Astronomy, Shanghai Jiao Tong University, Shanghai 200240, China}

\author{Mingpu Qin} \thanks{qinmingpu@sjtu.edu.cn}
\affiliation{Key Laboratory of Artificial Structures and Quantum Control (Ministry of Education),  School of Physics and Astronomy, Shanghai Jiao Tong University, Shanghai 200240, China}

\affiliation{Hefei National Laboratory, Hefei 230088, China}

\date{\today}


\begin{abstract}  
Classical and quantum states can be distinguished by entanglement entropy, which can be viewed as a measure of quantum resources. Entanglement entropy also plays a pivotal role in understanding computational complexity in simulating quantum systems. However, stabilizer states formed solely by Clifford gates can be efficiently simulated with the tableau algorithm according to the Gottesman-Knill theorem, although they can host large entanglement entropy. In this work, we introduce the concept of non-stabilizerness entanglement entropy which is basically the minimum residual entanglement entropy for a quantum state by excluding the contribution from Clifford circuits. It can serve as a new practical and better measure of difficulty in the classical simulation of quantum many-body systems. We discuss why it is a better criterion than previously proposed metrics such as Stabilizer R\'enyi Entropy. We also show numerical results of non-stabilizerness entanglement entropy with concrete quantum many-body models. The concept of non-stabilizerness entanglement entropy expands our understanding of the ``hardness'' in the classical simulation of quantum many-body systems.      

\end{abstract}

\maketitle

\section{Introduction}

The pursuit of quantum ``supremacy'' or quantum advantage is intricately related to the simulation of quantum many-body systems. Fundamentally, quantum computing can be viewed as a sophisticated manipulation of an extensive quantum many-body system. One of the most promising applications for quantum computers is solving quantum many-body problems that remain difficult for classical computers~\cite{feynman2018simulating}, such as uncovering the mechanisms behind high-temperature superconductivity~\cite{qin2022hubbard,annurev-conmatphys-031620-102024}. However, it is important to acknowledge that not all simulations of quantum many-body systems can be used to demonstrate quantum advantage, or not all quantum circuits present a challenge for classical computation, even though their corresponding Hilbert spaces grow exponentially with the system size. On one hand, physicists have developed several highly effective classical algorithms which are efficient for special quantum systems—such as dynamical mean-field theory~\cite{RevModPhys.68.13, RevModPhys.78.865}, quantum Monte Carlo methods~\cite{RevModPhys.87.1067,RevModPhys.73.33}, tensor network techniques~\cite{RevModPhys.93.045003,xiang2023density} and so on (see~\cite{PhysRevX.5.041041} for a review). On the other hand, certain specialized quantum circuits, including Clifford~\cite{gottesman1998heisenberg,Nielsen_Chuang_2010} and Matchgate circuits~\cite{10.1137/S0097539700377025}, can be efficiently simulated on classical computers with polynomial complexity using specific methodologies~\cite{PhysRevA.70.052328,PhysRevA.73.022334,doi:10.1098/rspa.2008.0189,PhysRevA.65.032325}.

It is arguably believed that the classical simulation of general quantum many-body systems could be difficult. Recently, with the rapid development of hardware in quantum computing, people try to demonstrate the quantum advantages with these devices. 
Google first claimed the quantum ``supremacy'' in the sampling of random quantum circuits, a task takes 200 seconds in the 53-qubits Sycamore processor which they claimed requires a classical computer to run 10,000 years to reproduce the results~\cite{Google_Nature}. However, subsequent analyses of the experimental quantum circuits revealed underlying structures that allowed these results to be replicated on a classical computer in just a few hundred seconds~\cite{pednault2019leveragingsecondarystoragesimulate,Gray2021hyperoptimized,pan2021simulatingsycamorequantumsupremacy,PhysRevLett.129.090502,10.1145/3458817.3487399,PRXQuantum.4.020304}. Similarly, another quantum super-experiment~\cite{doi:10.1126/science.abe8770} has also been debated in comparison with some classical simulations~\cite{popova2022crackingquantumadvantagethreshold,villalonga2022efficientapproximationexperimentalgaussian,PhysRevLett.128.190501}. Recently, IBM built a 127-qubits quantum computer and claimed it had the ability to simulate the dynamics of the kicked Ising model, while achieving more accurate results than classical simulations~\cite{kim2023evidence}. However, subsequent classical simulations of the IBM experiment indicated that the era of quantum "supremacy" has not yet fully arrived~\cite{liao2023simulationibmskickedising,doi:10.1126/sciadv.adk4321,PRXQuantum.5.010308}. It “seems” possible to find ways to make classical simulation for a quantum many-body systems faster, once we get some characterization of the corresponding Hilbert space physical structure. These debates raise critical questions about which types of quantum many-body systems are genuinely difficult to simulate, what results truly demonstrate quantum advantage, and whether the difficulty of simulating a quantum state can be quantitatively assessed. These questions are intricately associated with the measure of the complexity of simulation a quantum many-body systems in a classical computer.

Over the past few decades, several concepts have been proposed to assess the complexity of classically simulating quantum many-body systems. Entanglement entropy is a well known metric, with the understanding that higher entanglement entropy indicates greater difficulty in classical simulation~\cite{PhysRevLett.78.2275,RevModPhys.81.865}. Following the Gottesman-Knill theorem~\cite{gottesman1998heisenberg,Nielsen_Chuang_2010}, it was recognized that the number of non-Clifford gate operations needed also serves as an indicator of simulation difficulty~\cite{PhysRevA.71.022316,Veitch_2014,PhysRevX.6.021043,PhysRevLett.118.090501,howard2014contextuality, PRXQuantum.2.010345}, since Clifford gate operations can be efficiently simulated on classical computers. This measure is also referred to as non-stabilizerness or quantum magic. We now know that stabilizer states formed solely by Clifford gates can support high entanglement~\cite{PhysRevX.7.031016}, but they can be efficiently simulated on classical computers according to the Gottesman-Knill theorem. However, we can also find examples with zero entanglement but non-zero magic. As we will discuss later, we can construct product states with non-zero magic. These discrepancies challenge the current metrics used to measure the complexity of simulating a quantum state and highlight the need for a more nuanced understanding of the factors that contribute to the complexity of simulating a quantum many-body systems.

Moreover, calculating the magic of a quantum system is generally a hard problem~\cite{Veitch_2014,PhysRevX.6.021043,PhysRevLett.116.250501,PhysRevLett.115.070501,PRXQuantum.3.020333,Bravyi2019simulationofquantum,Heinrich2019robustnessofmagic,vandewetering2024optimisingquantumcircuitsgenerally}. Although numerous effective computational methods have been developed~\cite{PhysRevLett.115.070501,Bravyi2019simulationofquantum,PhysRevLett.131.180401,PhysRevLett.133.010601}, they are primarily applicable only to small system sizes or special systems. 

In this work, we introduce a new metric termed Non-stabilizerness Entanglement Entropy (NsEE), which integrates entanglement entropy and quantum magic. We show NsEE is a better metric to characterize the difficulty in the classical simulation of quantum many-body systems. We also show that NsEE can be efficiently calculated in the framework of Clifford circuits Augmented Matrix Product States (CA-MPS)~\cite{qian2024augmentingdensitymatrixrenormalization}. With CA-MPS, we calculate the NsEE for the ground state of Toric code model, 2D transverse Ising model, 2D XXZ model and also for random quantum circuits. CA-MPS gives a zero NsEE for the ground state of Toric code model as expected, demonstrating the effectiveness of this method. 

The rest of the article is organized as follows. In Sec.~\ref{Preliminaries}, we introduce some essential preliminaries for subsequent discussions. Specifically, in Sec.~\ref{EE_TN} we introduce the connection between Entanglement Entropy and Tensor Network methods. In Sec.~\ref{Stab_state}, we introduce the definition of stabilizer state and give a proof of the flatness of the entanglement spectrum for stabilizer state. In Sec.~\ref{Stab_tf}, we explain how stabilizer state evolution under Clifford circuits can be efficiently simulated according to the Gottesman-Knill theorem. In Sec.~\ref{CA-MPS_section}, we give an overview of the CA-MPS method we developed before. In Sec.~\ref{SRE_Sec}, we briefly review Stabilizer R\'enyi Entropy (SRE). In Sec.~\ref{NsEE_sec}, we introduce the definition of Non-stabilizerness Entanglement Entropy. In Sec.~\ref{Numerical_Sec}, we show the numerical results of NsEE for the ground state of several models and discuss the practical utility of NsEE. We conclude our work and discuss the perspective of NsEE in Sec.~\ref{Con_Per_Sec}.

\section{Preliminaries}
\label{Preliminaries}

Before introducing the concept of Non-stabilizerness Entanglement Entropy, we first provide an overview of the essential preliminaries. 

\subsection{Entanglement Entropy and Tensor Network methods}
\label{EE_TN}
Entanglement Entropy (EE) is a measure of the quantum correlations between different parts of a quantum system.
In the field of tensor networks~\cite{xiang2023density,RevModPhys.93.045003}, the entanglement entropy of a quantum state plays a crucial role in determining the computational cost required for its accurate simulation. 
For a system with density matrix $\rho = |\psi\rangle\langle\psi|$ ($|\psi\rangle$ represents a pure state),
by dividing this system into parts $A$ and $B$, the corresponding entanglement entropy $S_{A(B)}$ of part $A(B)$ is defined as:
\begin{equation}
    S_A = -\operatorname{Tr}(\rho_A \log \rho_A) = -\operatorname{Tr}(\rho_B \log\rho_B) = S_B
    \label{EE_eq}
\end{equation}
where $\rho_{A(B)} = \operatorname{Tr}_{B(A)}(\rho)$ is reduced density matrix of A(B) part, obtained by performing a partial trace of the total density matrix $\rho$ over the Hilbert space of the other part. For a pure state $|\psi\rangle$, we can rewrite it as $|\psi\rangle = \sum_{i} \lambda_i |i_A\rangle|i_B\rangle$, where $|i_A\rangle$ and $|i_B\rangle$ are the Schmidt basis of the reduced density matrix $\rho_A$ and $\rho_B$, respectively. Hence the entanglement entropy can be expressed as:
\begin{equation}
    S_A = -\sum_i \lambda_i^2 \log \lambda_i^2
    \label{EE_eq2}
\end{equation}

This is closely related to the Singular Value Decomposition (SVD) process in tensor networks~\cite{PhysRevLett.69.2863,PhysRevLett.91.147902,RevModPhys.77.259,SCHOLLWOCK201196}, where $\lambda_i$ represents the singular values obtained from the SVD. For example, the bond dimension of the Matrix Product States (MPS) is bounded by the number of nonzero $\lambda_i$~\cite{SCHOLLWOCK201196}. Thus, the entanglement entropy of a state that can be represented by an MPS is bounded by $\log(D)$, where $D$ is the bond dimension of the MPS. As a result, to accurately or faithfully represent a state, larger entanglement entropy generally requires a larger bond dimension, which increases the needed computational resources in the simulation.

\subsection{Stabilizer state}
\label{Stab_state}

Let us consider a quantum system composed of $N$ qubits. Each qubit has a Hilbert space of dimension 2, typically denoted as $\mathbb{C}^2$. Therefore, the whole Hilbert space of the entire system is the tensor product of the Hilbert spaces of the individual qubit:
$\mathcal{H} = (\mathbb{C}^2)^{\otimes N}$.
The dimension of this total Hilbert space is $2^N$. 
The stabilizer group $\mathcal{G}$ is defined to be the Abelian group generated by the following stabilizer generators $\{g_1, g_2, \ldots, g_N\}$ acting on $\mathcal{H}$. Each $g_i$ is a product of Pauli operators. In other words, $g_i$ is a product of operators acting on individual sites. And on each site, the operator (up to a phase factor $\{\pm 1, \pm i\}$) can be chosen from the set $\{I, X, Y, Z\}$. By definition, we have $g_i^2 = 1$, and each $g_i$ has eigenvalues $\pm 1$. The number of the group element is $|\mathcal{G}| = 2^N$. It is known that there exists a unique state that is determined by the stabilizer generators~\cite{gottesman1998heisenberg,Nielsen_Chuang_2010}. 

The stabilizer state is defined as a state that satisfies:
\begin{equation}
g_i|\psi\rangle = |\psi\rangle \quad \forall i \in \{1, 2, 3,\cdots, N\}
\Leftrightarrow  h|\psi\rangle = |\psi\rangle \quad \forall h \in \mathcal{G}
\end{equation}
The stabilizer state can be prepared by a Clifford circuit $\mathcal{C}$, which only consists of Clifford gates: $\{ \text{Hadamard}, S, \text{CNOT} \}$, starting from the state $|0\rangle^{\otimes N}$ ($|0\rangle$ is the eigenstate of $Z$ with eigenvalue 1)~\cite{gottesman1998heisenberg,Nielsen_Chuang_2010}. In other words, the stabilizer state can also be expressed as:
$|\psi\rangle = \mathcal{C}|0\rangle^{\otimes N}.$

The density matrix $\rho$ of this state can then be written as
\begin{equation} 
    \begin{split}
    \rho &=  |\psi\rangle \langle \psi| \\
    &= \mathcal{C}(|0\rangle^{\otimes N}\langle 0|^{\otimes N})\mathcal{C}^{\dagger} \\
    &= \mathcal{C}(\prod_{i=1}^{N}\frac{1+Z_i}{2})\mathcal{C}^{\dagger} \\
    &= \frac{1}{2^N}\prod_{i=1}^{N}(1+g_i)\\
    &= \frac{1}{2^N}\sum_{h \in \mathcal{G}}h\\
    \end{split}
\end{equation}
where in the fourth line, we have used the fact that $\mathcal{C}Z_i\mathcal{C}^{\dagger}$ are the stabilizer generators for $|\psi\rangle = \mathcal{C}|0\rangle^{\otimes N}$.

To calculate the reduced density matrix for $\rho$, let us consider a subsystem $A (B)$ of the total system with $N_A (N_B)$ representing the number of qubits in $A (B)$. Define $\mathcal{G}_A$ as the subgroup of $\mathcal{G}$ associated with $A$: $\mathcal{G}_A \equiv \{ h_A \mid h_A \otimes \mathbb{I}_B \in \mathcal{G} \}$. 

The reduced density matrix of the stabilizer state $|\psi\rangle$ on subsystem $A$ can then be written as: 
\begin{equation}
    \begin{split}
    \rho_A &= \operatorname{Tr}_B \left( |\psi\rangle \langle\psi| \right) \\
    &= \frac{\text{Tr}(\mathbb{I}_B)}{2^N}\sum_{h_A \in \mathcal{G}_A} h_A \\
    &= \frac{2^{N_B}}{2^N}\sum_{h_A \in \mathcal{G}_A} h_A\\
    &= \frac{1}{2^{N_A}} \sum_{h_A \in \mathcal{G}_A} h_A\\
    \end{split}
\end{equation}
Given the fact that $\rho_A^2 = |\mathcal{G}_A|/2^{N_A}\rho_A$. All the non-zero eigenvalues of $\rho_A$ are the same: $\lambda= |\mathcal{G}_A|/2^{N_A}$. So the non-zero entanglement spectrum of stabilizer state is perfectly flat. In the later discussion, we will show numerical results to confirm the flatness of the entanglement spectrum of stabilizer state. It should also be noticed that if we expand the stabilizer state in the computational basis, the absolute value of the non-zero coefficients are also exactly the same. This is consistent with the fact that stabilizer state is special and to reach a universal entanglement spectrum, we must add non-Clifford operations to the Clifford circuits~\cite{10.21468/SciPostPhys.9.6.087,OLIVIERO2021127721,True2022transitionsin,PhysRevB.107.134202,PhysRevA.109.L040401}. In the following subsection, we will discuss how to efficiently simulate the stabilizer state classically.

\subsection{Clifford circuits}
\label{Stab_tf}
In classical computation, one can define a universal set of logic gate operations such as $\{$AND, NOT, OR$\}$ that can be used to perform any boolean function. A similar analog in quantum computation is to have a set of quantum gates that can approximate any unitary transformation up to the desired accuracy. One such universal quantum gate set is the  
Clifford + T set: $\{$H, S, CNOT, T$\}$, where the gates H, S, and CNOT are the generators of the Clifford group. The elements of this group are called Clifford gates, which transform Pauli string to Pauli string under conjugation. This means an  n-qubit unitary  $C$ belongs to the Clifford group if the conjugates  $CPC^\dagger$ are also Pauli string for all n-qubit Pauli sring $P=\sigma_1 \otimes \sigma_2\cdots \otimes \sigma_N$ ($\sigma_i\in \{I, \sigma^x, \sigma^y, \sigma^z\}$, $\sigma^{\alpha}$ is Pauli matrix). This can be checked by conjugating the Pauli matrix operation with the elements of the Clifford set defined above as shown in Table.~\ref{Pauli evolution}. The quantum circuits that consist only of Clifford gates are called Clifford circuits. 

A key property of Clifford circuits is that the evolution of stabilizer states under Clifford circuits can be efficiently simulated on classical computers, as established by the Gottesman-Knill theorem~\cite{gottesman1998heisenberg,Nielsen_Chuang_2010}, using the stabilizer tableau formalism~\cite{PhysRevA.70.052328}. As mentioned earlier, specifying a stabilizer state requires only the identification of its associated stabilizer group. This means we can effectively monitor the evolution of the state by tracking the generators of the stabilizer group throughout the Clifford evolution.

\begin{table}
    \centering
    \begin{tabular}{c|c|c} 
    \toprule
    Clifford Gate ($U$) & Input ($g$) & Output ($UgU^\dagger$) \\
    \cline{2-3} 
    \hline
    \multirow{2}{*}{$\text{Hadamard}$} & $X$ & $Z$ \\
    & $Z$ & $X$ \\
    \hline
    \multirow{4}{*}{$\text{CNOT}_{12}$} & $X_1$ & $X_1X_2$ \\
    & $X_2$ & $X_2$ \\
    & $Z_1$ & $Z_1$ \\
    & $Z_2$ & $Z_1Z_2$ \\
    \hline
    \multirow{2}{*}{$\text{S}$} & $X$ & $Y$ \\
    & $Z$ & $Z$ \\
    \hline
    \end{tabular}
    \caption{The transformation of Pauli operator under Clifford gates.}
    \label{Pauli evolution}
\end{table}

As an example, consider a stabilizer state $|\psi\rangle$ with  stabilizer group generators $S = \{g_1, g_2, \ldots, g_N\}$. After applying Clifford circuits $U$ to this state, we can get the new state $U|\psi\rangle$ with new stabilizer group generators:
\begin{equation}
    USU^\dagger = U\{g_1, g_2, \ldots, g_N\}U^\dagger
\end{equation}
Since $g_i$ is stabilizer of $|\psi\rangle$, we have $Ug_iU^\dagger U|\psi\rangle =  U|\psi\rangle$. We need to notice that the Clifford circuits never change the number of the generators, the evolution of a Pauli string under Clifford circuits results in another Pauli string. By utilizing the stabilizer tableau algorithms, the complexity of tracking the evolution of generators under Clifford circuits is $O(N^2)$~\cite{PhysRevA.70.052328}. In contrast, directly tracking the evolution of a stabilizer state under a Clifford circuit has a complexity exponential in $N$.

\subsection{Clifford circuits Augmented Matrix Product States}
\label{CA-MPS_section}
The CA-MPS method is designed to enhance the capabilities of classical simulations by leveraging the advantages of both tensor network and stabilizer formalism~\cite{qian2024augmentingdensitymatrixrenormalization}.

MPS is defined as:
\begin{equation}
    |\text{MPS}\rangle = \sum_{\{\sigma_i\}} \text{Tr}(A^{\sigma_1}_1A^{\sigma_2}_2\cdots A^{\sigma_N}_N)|\sigma_1\sigma_2\cdots \sigma_N\rangle
    \label{MPS}
\end{equation}
where $A$ is a rank-3 tensor with a physical index $\sigma_i$ (with dimension $2$ for spin $1/2$ degree of freedom) and two auxiliary indices with dimension $D$. As we mentioned above, the entanglement entropy of MPS is bounded by $\log(D)$.

We can apply Clifford circuits to the MPS wave function to reduce the entanglement entropy. The new wave function, which we name as CA-MPS, is defined as:
\begin{equation}
    |\text{CA-MPS}\rangle = \mathcal{C}|\text{MPS}\rangle
    \label{CA-MPS}
\end{equation}
where $\mathcal{C}$ denotes the Clifford circuits. Accordingly, we need to transform the physical observables using the same Clifford circuits. For example, the ground state energy can be written as: $\langle\text{MPS}|H_{\text{MPS}}|\text{MPS}\rangle=\langle\text{MPS}|\mathcal{C}^{\dagger}\mathcal{C}H_{\text{MPS}}\mathcal{C}^{\dagger}\mathcal{C}|\text{MPS}\rangle=\langle\text{CA-MPS}|H_{\text{CA-MPS}}|\text{CA-MPS}\rangle$ with $H_{\text{CA-MPS}}=\mathcal{C}H_{\text{MPS}}\mathcal{C}^{\dagger}$. The Hamiltonian and other physical observables can be usually represented as a summation of Pauli strings. And we know that Clifford circuits preserve the structure of Pauli string. So the number of Pauli string terms in $H_{\text{CA-MPS}}$ remains the same as in $H_{\text{MPS}}$. Otherwise, for general unitary transformation, the number of terms in the transformed Hamiltonian could explode, making the calculation infeasible.

 By carefully selecting appropriate Clifford transformations, the resulting wave function can exhibit reduced entanglement entropy compared to the original wave function. This reduction in entanglement entropy can significantly enhance the efficiency and performance of classical simulations. The CA-MPS method offers an efficient classical simulation technique for identifying the optimal Clifford transformation, which effectively disentangles the original wave function. 

\begin{figure}[t]
    \includegraphics[width=80mm]{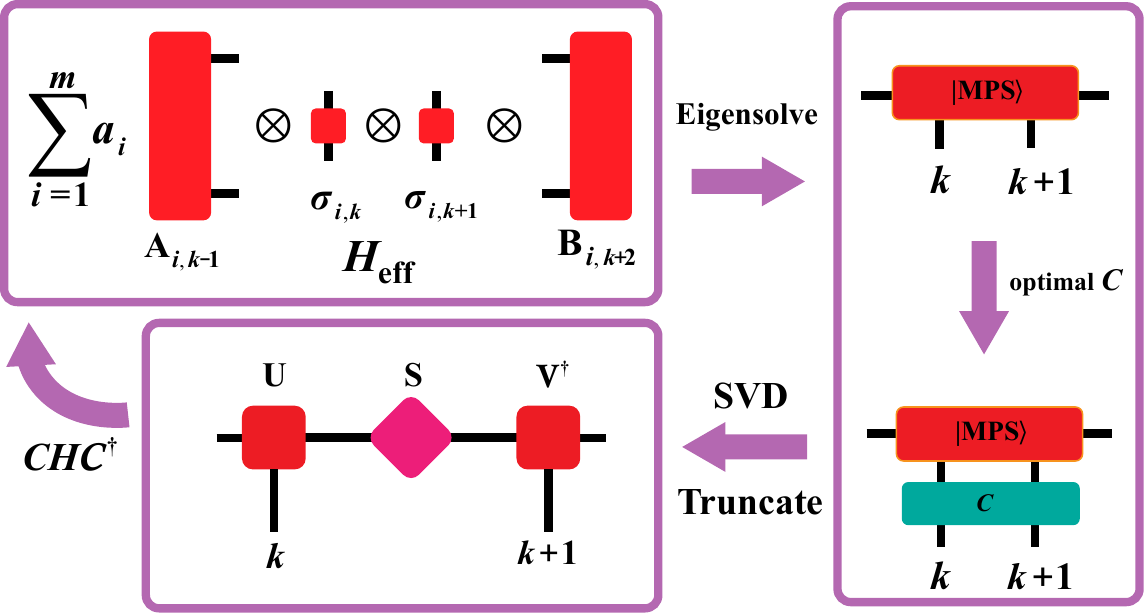}
       \caption{Schematic illustration of CA-MPS method. CA-MPS introduces a tiny modification to the DMRG algorithm. First, we calculate the ground state of $H_{\text{eff}}$ using the DMRG method. Then before the truncation, we apply the optimal two-qubit Clifford circuits to the MPS state aiming to minimize the truncation error. We can iterate all the possible two-qubit Clifford circuits~\cite{10.1063/1.4903507,PhysRevB.100.134306,PhysRevA.87.030301} to find the optimal one, i.e., giving the smallest truncation error. We also need to transform the original Hamiltonian to $CHC^\dagger$ using the obtained optimal Clifford circuits. Then we move to the next sites and repeat the process. The details can be found in~\cite{qian2024augmentingdensitymatrixrenormalization}.}
       \label{CA-MPS_Fig}
\end{figure}

A schematic illustration of the CA-MPS algorithm is shown in Fig.~\ref{CA-MPS_Fig}. The algorithm introduces a tiny modification to the DMRG method~\cite{PhysRevLett.69.2863}. In one step of the CA-MPS optimization process, we first calculate the ground state $|\text{MPS}\rangle$ of the effective Hamiltonian $H_{\text{eff}}$ for sites $k$ and $k+1$ with DMRG. Instead of directly applying a truncated SVD to the $|\text{MPS}\rangle$, we employ a two-qubit Clifford circuit $C$ to the $|\text{MPS}\rangle$ to minimize the discarded weight during truncation. The optimal two-qubit Clifford circuits can be determined by iterating all the possible two-qubit Clifford circuits~\cite{10.1063/1.4903507,PhysRevB.100.134306,PhysRevA.87.030301}. Given that $CH_{\text{eff}}C^{\dagger}C|\text{MPS}\rangle = E_gC|\text{MPS}\rangle$, we also need to transform the original Hamiltonian to $H' = CHC^{\dagger}$. Subsequently, same as in DMRG, we shift from sites $k$ and $k+1$ to sites $k+1$ and $k+2$, continuing the optimization process until all sites are optimized. This constitutes a CA-MPS sweep. The sweeping is repeated until the entanglement entropy and energy converges.

CA-MPS has been successfully applied to the simulation of quantum many-body systems. Not only can it efficiently calculate the ground state of a quantum system~\cite{qian2024augmentingdensitymatrixrenormalization}, but it can be also extended to time evolution~\cite{qian2024cliffordcircuitsaugmentedtimedependent, mello2024clifforddressedtimedependentvariational}. In the ground state calculation, the CA-MPS method can reduce the entanglement entropy of the ground state by a factor of $30\%$ to $50\%$~\cite{qian2024augmentingdensitymatrixrenormalization}. For special cases, such as the Toric code model with a stabilizer state as the ground state, the entanglement entropy can be reduced to exactly zero as we will show in later discussions.

Since the magic of a quantum state is invariant under the evolution of Clifford circuits, and the CA-MPS wave function always gives a lower bond dimension than the original MPS wave function, the CA-MPS method can also benefit the calculation of other magic metrics of a quantum state, such as the Stabilizer R\'enyi Entropy which will be discussed in the following subsection.

\subsection{Stabilizer R\'enyi Entropy}
\label{SRE_Sec}
Determining whether a quantum state is easy or hard to simulate on a classical computer is a fundamental question in the fields of quantum many-body physics and quantum information theory. Entanglement entropy often plays a pivotal role in addressing this question. However, as demonstrated by the Gottesman-Knill theorem, stabilizer states under Clifford circuits evolution can be efficiently simulated on classical computers despite possessing substantial entanglement entropy. Within the context of stabilizer states, another quantity becomes significant: the measure of non-Clifford operations cost on a quantum state. This is crucial because Clifford operations are effectively cheap in classical simulations. The quantity, known as non-stabilizerness or quantum magic, quantifies the computational overhead introduced by non-Clifford operations.

Nevertheless, measuring the non-stabilizerness of a quantum state presents a significant challenge. Various metrics have been developed for this purpose, including stabilizer extent~\cite{Bravyi2019simulationofquantum}, stabilizer fidelity~\cite{Bravyi2019simulationofquantum}, stabilizer rank~\cite{PhysRevX.6.021043,Bravyi2019simulationofquantum,PhysRevLett.116.250501}, Wigner negativity and mana~\cite{Veitch_2014,PhysRevLett.115.070501}. Although these measures are well-defined theoretically, their practical computation often proves difficult.

The Stabilizer R\'enyi Entropy (SRE)~\cite{PhysRevLett.128.050402} was recently introduced as a measure for quantifying the non-stabilizerness of a quantum state. Calculating the exact value of SRE for a general quantum state is also hard, as it requires the expectation values of all Pauli strings (see the definition below in Eq.~(\ref{SRE_def})). However, efficient approximation methods have been developed to facilitate its practical computation. Techniques such as employing MPS for sampling~\cite{Haug2023stabilizerentropies, PhysRevLett.131.180401} and using Pauli basis transformations with compressed bond dimensions~\cite{PhysRevB.107.035148, PhysRevLett.133.010601} enable these approximations, making SRE an accessible measure for quantifying non-stabilizerness in quantum states. Additionally, the concept of SRE can be extended to the Heisenberg picture, leading to the operator stabilizer entropy~\cite{dowling2024magicheisenbergpicture}.

There was debate regarding whether the SRE qualifies as a magic monotone before. Specifically, it was shown that for orders $ n < 2 $, the SRE does not serve as a magic monotone~\cite{Haug2023stabilizerentropies}. However, subsequent research established that for $ n \geq 2 $, the SRE indeed behaves as a magic monotone~\cite{leone2024stabilizerentropiesmonotonesmagicstate}, confirming its utility in measuring the magic of quantum states.

For a pure state $|\psi\rangle$, the n-order SRE is defined as:
\begin{equation}
    M_n(\psi\rangle) = \frac{1}{1-n} \log \sum_{P\in P_N} \frac{\langle \psi|P|\psi\rangle^{2n}}{2^N}
    \label{SRE_def}
\end{equation}
where $P_N$ contains all possible Pauli string for $N$ qubits system.
As a magic resource measure, it has the following properties~\cite{PhysRevLett.128.050402}
\begin{enumerate}
\item $M_n \geq 0$ for any pure state and $M_n = 0$ for stabilizer state.
\item $M_n$ is invariant under Clifford operations. 
\item $M_n$ is additive, which means it grows extensively with the system size.
\end{enumerate}

However, this does not necessarily imply that SRE can be used to quantify the classical simulation complexity of a quantum state perfectly. For instance, simulating a product state is straightforward, yet the SRE of a product state is not necessarily zero. Consider the $T$ state $|T\rangle ^{\otimes N}$ with $ |T\rangle = \frac{ |0\rangle + e^{i\frac{\pi}{4}}|1\rangle }{\sqrt{2}}$, the 2-order SRE $M_2$ of this state is $ M_2 = -N\log[(1+\cos^4\frac{\pi}{4} + \sin^4\frac{\pi}{4})/2] > 0$. The entanglement entropy of this state is, however, exactly zero. This example demonstrates that to accurately quantify the classical simulation complexity of a quantum state, one must consider both non-stabilizerness and entanglement entropy.

\section{Non-stabilizerness Entanglement Entropy}
\label{NsEE_sec}
In this work, we introduce Non-stabilizerness Entanglement Entropy (NsEE), which is defined as the minimum entanglement entropy (EE) achievable, by applying Clifford circuits to the studied state. From the definition, we know that NsEE is a metric to measure the difficulty of classical simulation of quantum many-body systems by taking both non-stabilizerness and entanglement entropy into consideration.

\begin{figure}[t]
    \includegraphics[width=80mm]{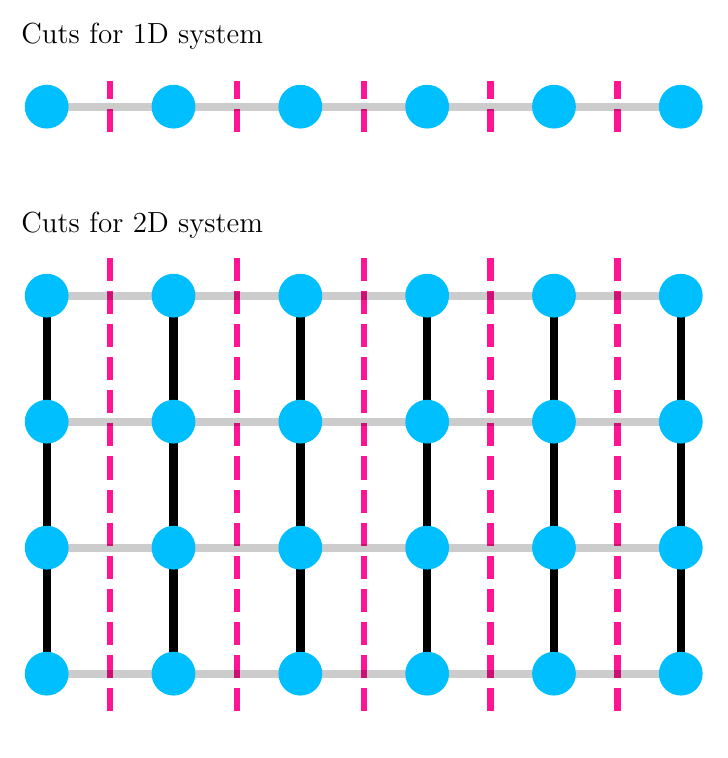}
       \caption{Defined bipartitions (cuts) for calculating NsEE in both 1D and 2D systems with open boundary conditions. Similar bipartitions can be also defined for systems under periodic boundary conditions. }
       \label{EE_cuts}
\end{figure}

To calculate NsEE for a quantum state, we need to first specify the bipartition of the system.
In Fig.~\ref{EE_cuts}, without loss of generality, we provide examples of how to define the bipartition (cut) for both 1D and 2D systems, which can be straightforwardly generalized to higher-dimensional systems.
NsEE is calculated by considering the EE for each defined cut and then summing over all the resulting EEs. Mathematically, this can be expressed as:
\begin{equation}
    \text{NsEE}({|\psi\rangle})= \mathop{\min}_{\{\mathcal{C}\}}\sum_{\text{cuts}} \text{EE}({\mathcal{C}|\psi\rangle}).
    \label{NsEE}
\end{equation} 

This definition encompasses many key requirements for both magic resource measure and entanglement measure.
\begin{enumerate}
\item NsEE $\geq$ 0 for any pure state.
\item NsEE is zero for stabilizer states, $\text{NsEE}({|Stab.\rangle})=0$.
\item Stability under Clifford operations, i.e., $\text{NsEE}({\mathcal{C}|\psi\rangle})=\text{NsEE}({|\psi\rangle})$.
\item NsEE is also additive, $\text{NsEE}({|\psi\rangle_A \otimes |\psi\rangle_B})=\text{NsEE}({|\psi\rangle_A})+\text{NsEE}({|\psi\rangle_B})$.
\end{enumerate}

\begin{figure}[t]
    \centering
    \includegraphics[width=80mm]{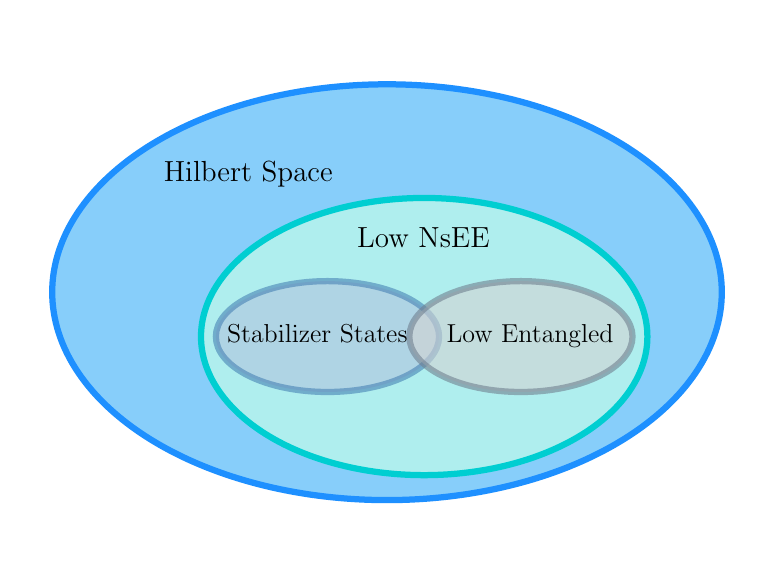}
    \caption{Schematic illustration of the relationship among low NsEE state, stabilizer state, and low entangled state in quantum many-body Hilbert space.}
    \label{Hilbert}
\end{figure}

If the additive condition is released, we can define NsEE by the vertical cut which separates the system into two equal parts. 

NsEE exhibits additional properties that establish it as a more effective measure for assessing classical computational complexity. In certain instances, NsEE proves to be a more reliable indicator of classical computational complexity than SRE. Below, we outline several cases:
\begin{enumerate}
\item For state $|\psi\rangle$ represented as the product of T state: $|\psi\rangle=|T\rangle^{\otimes N}$, $M_2(|\psi\rangle) > 0 $, while $\text{NsEE}(|\psi\rangle)=0$.
\item Considering state $|\psi\rangle$ prepared by gate set $\{$S, CNOT, T$\}$, $M_2(|\psi\rangle)$ could be high, whereas $\text{NsEE}(|\psi\rangle)=0$.
\item For state $|\psi\rangle$  prepared by a Clifford + T circuit with limited number of T gates (approximately $N$):  $|\psi\rangle=\prod_{i=1}^{N_t}( T^{i}\otimes I^{\otimes N-1} U^{c_i})|0\rangle^{\otimes N}, \ N_t \lesssim N$, where $U^c$ denote the Clifford circuit,  $M_2(|\psi\rangle)$ is increase linearly against $N_ t$~\cite{haug2024probingquantumcomplexityuniversal}, while $\text{NsEE}(|\psi\rangle)$ remains very small.
\end{enumerate}

The first case is discussed in above sections. For the $T$ state $|T\rangle ^{\otimes N}$ with $ |T\rangle = \frac{ |0\rangle + e^{i\frac{\pi}{4}}|1\rangle }{\sqrt{2}}$, the 2-order SRE $M_2$ of this state is $ M_2 = -N\log[(1+\cos^4\frac{\pi}{4} + \sin^4\frac{\pi}{4})/2] > 0$. In the second case, we consider the state prepared by the gate set $\{$S, CNOT, T$\}$. Although the SRE for these states could be high, they can be efficiently simulated classically. This is based on the fact that CNOT, T, and S gates commute with each other. Thus, for any state that is prepared by $\{$S, CNOT, T$\}$ can be written as: $|\psi\rangle= \prod_{i=1}^{n} \text{CNOT}^{i}|s\rangle^{\otimes N}$, where $|s\rangle$ is single qubit state. However, the SRE of such states is not necessarily zero, while NsEE is obviously zero.

We would also like to emphasize that Clifford + T circuits, without constraint on the number of T gates, are universal for quantum state preparation. The third example demonstrates how NsEE effectively captures the complexities of classical simulations by monitoring the transition from easy to hard as the number of T gates increases (additional discussions can be found in the subsequent sections). 

In Fig.~\ref{Hilbert}, we illustrate how NsEE reflects the classical computational complexity associated with both the stabilizer formalism and the tensor network formalism. Specifically, stabilizer states can be efficiently simulated on classical computers by the Gottsman-Knill theorem, while low-entangled states can be effectively simulated using tensor network methods.  The NsEE effectively captures both of these characteristics in classical computations.

Finally, having demonstrated the effectiveness of the NsEE in capturing classical computational complexity, we now summarize a possible procedure for calculating the NsEE for a given state $|\psi\rangle$ :
\begin{enumerate}
\item Defining the specific bipartitions of the system into two subsystems as shown in Fig.~\ref{EE_cuts}.
\item Applying various Clifford circuits $\mathcal{C}$ to the state $|\psi\rangle$.
\item Calculating the EE for each Clifford-transformed state with respect to the chosen bipartitions.
\item Finding the minimum EE among all Clifford-transformed states and summing the EEs for each cut to obtain NsEE.
\end{enumerate}

Strictly speaking, it is nearly impossible to precisely calculate the NsEE for a quantum state. This is due not only to the notorious difficulty in the computation of the EE but also because the number of independent Clifford transformations grows exponentially with the system size. However, the recently developed CA-MPS method~\cite{qian2024augmentingdensitymatrixrenormalization} provides an effective tool to calculate NsEE. With CA-MPS, we can not only efficiently calculate the entanglement entropy but also determine the minimum entanglement entropy after Clifford transformations (more discussions about the CA-MPS method are provided in the preliminary Sec.~\ref{CA-MPS_section}).

\begin{figure}[t]
    \includegraphics[width=80mm]{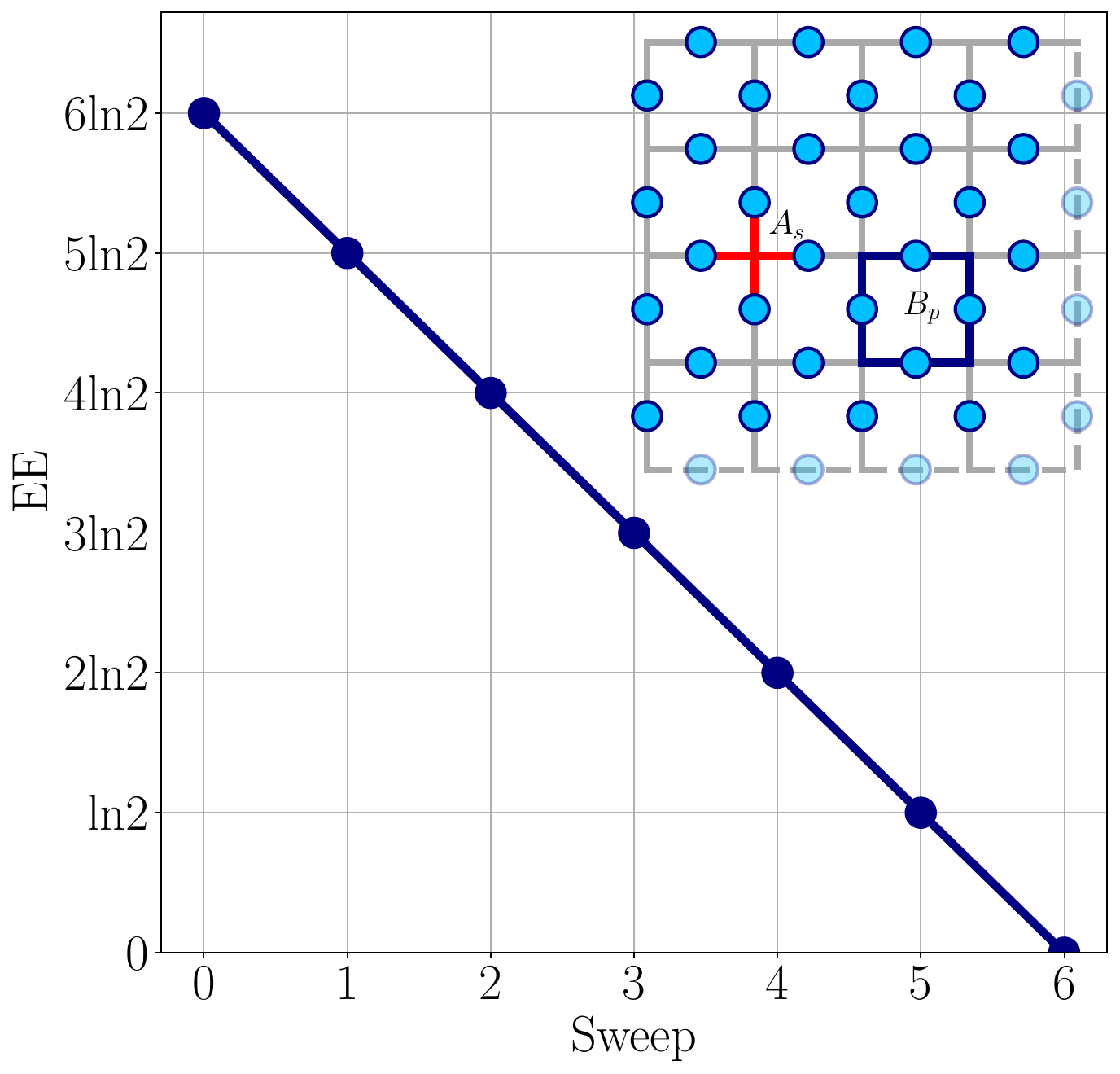}
       \caption{The maximum entanglement entropy (EE) in the ground state of Toric code mode among all cuts as a function of the CA-MPS sweep. The calculated system, a $4 \times 8$ lattice under periodic boundary conditions, is shown in the inset.}
       \label{4Toric}
\end{figure}

\section{Numerical results}
\label{Numerical_Sec}
In this section, we examine specific models, including the Toric code model, 2D Transverse Ising model, 2D XXZ model, and random circuits to illustrate the practical utility of NsEE and to highlight some of its properties.

\subsection{Toric code model}
First, we demonstrate that the CA-MPS algorithm can efficiently identify the associated Clifford transformation that disentangles a stabilizer state into a product state. A notable example of this is the well-known Toric code model, whose ground state is a stabilizer state~\cite{KITAEV20032}. The Hamiltonian of the toric
code reads:
\begin{equation}
    H_{\text{Toric}} = -\sum_{s}A_s - \sum_{p}B_p
    \label{H_Toric}
\end{equation}
where  $A_s = \prod_{i\in s} \sigma^x_i$ and $B_p = \prod_{i\in p} \sigma^z_i$ ($\sigma^{\alpha}_i$ are the usual Pauli matrix).
Subscript $s$ ($p$) refers to sites (plaquettes)
of a square lattice and $i$ runs over all bonds where spins
are located as shown in the inset of Fig.~\ref{4Toric}. Because $[A_s, B_p] = 0$ for any $s$ and $p$, and $A_s$ and $B_p$ are both Pauli strings, the ground state of Eq.~(\ref{H_Toric}) is the stabilizer state determined by $A_s$ and $B_p$.   

The simulation results using CA-MPS method are presented in Fig.~\ref{4Toric}, where we consider a $4\times8$ lattice with periodic boundary conditions (PBC). The NsEE is slightly different from the definition in Eq.~(\ref{NsEE}). Here we consider the cut that gives the maximum entanglement entropy, i.e., the cut that divides the system into two equal parts. The entanglement entropy of the ground state for a system with width $L$ is given by: $\text{EE}=(2L-2) \text{ln}(2)$~\cite{hermanns2017entanglementtopologicalsystems}. 
From Fig.~\ref{4Toric}, we observe that the entanglement entropy decreases by $\text{ln}(2)$ during each optimization sweep in CA-MPS, and ultimately approaches zero after several steps.
This suggests that for a system of width $L$, CA-MPS can efficiently identify the corresponding Clifford transformation in $2(L-2)$ sweeps for the Toric code model, thereby numerically verifying the Gottesman-Knill theorem, which states that stabilizer states can be efficiently learned on classical computers~\cite{pirsa_PIRSA:08080052}. Consequently, the corresponding NsEE and SRE for such a state are both zero, consistent with the inherent properties of each definition.

\subsection{2D transverse Ising model}

The 2D transverse Ising model is a fundamental model in quantum statistical mechanics and condensed matter physics, particularly useful for studying quantum phase transitions. 
The Hamiltonian of the transverse Ising model is defined as:
\begin{equation}
    H_{\text{Ising}} = -J\sum_{\langle i,j \rangle}\sigma^z_i \sigma^z_j - h\sum_i \sigma^x_i
    \label{H_Ising}
\end{equation}
where $\sigma^x$ and $\sigma^z$ are the Pauli operators represented in the $x$ and $z$ direction at site $i$. $h$ is the strength of the transverse magnetic field, and the coupling strength $J$ is set to 1 as the unit of energy. $\langle i,j \rangle$ denotes summation over nearest-neighbor pairs. We consider open boundary conditions (OBC) on a 2D square lattice.

In the thermodynamic limit, the two-dimensional transverse Ising model undergoes a quantum phase transition from a ferromagnetic to a paramagnetic phase at approximately $h\approx 3.05$~\cite{PhysRevB.105.205102,PhysRevA.81.032304}. At this critical point, the EE exhibits significant changes~\cite{PhysRevA.81.032304}, making it a useful quantity for determining the critical point.

\begin{figure}[t]
    \includegraphics[width=85mm]{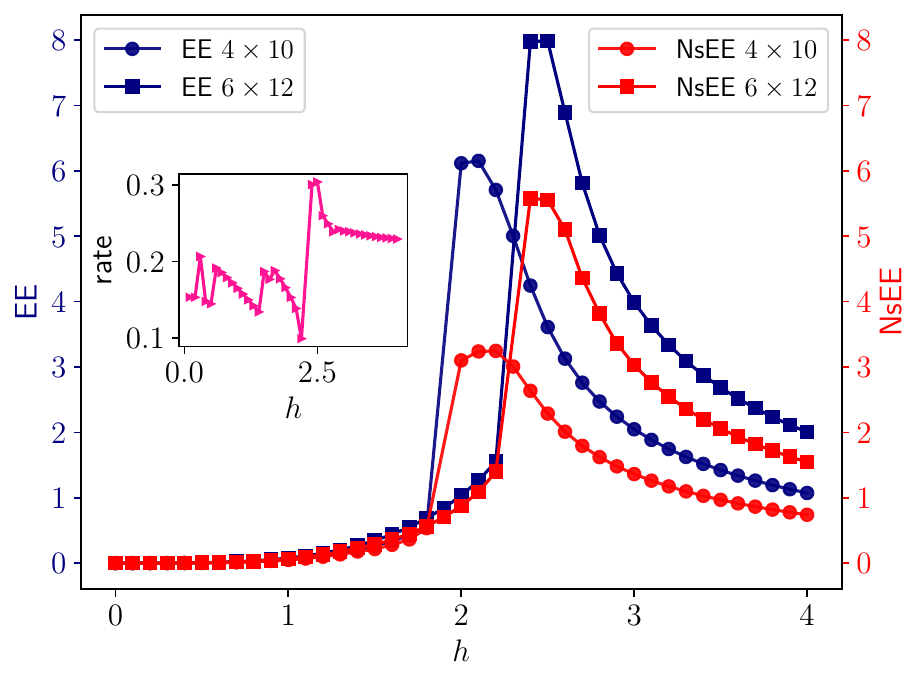}
       \caption{Numerical results for the 2D transverse Ising model with OBC. The entanglement entropy (EE) and the corresponding NsEE of the 2D cuts (Eq.~(\ref{NsEE})) as a function of the $h$ are shown. The insert shows the reduction rate between EE and NsEE for system size $6\times 12$.}
       \label{Ising}
\end{figure}

The results from CA-MPS are illustrated in Fig.~\ref{Ising} (Note: all EE and NsEE results are summed over all the different cuts as shown in Fig.~\ref{EE_cuts}). In the ferromagnetic phase at 
$h=0$, the ground state is a product state, leading to both EE and NsEE being zero. As $h$ increases, the ground state becomes increasingly complicated, and both EE and NsEE rise correspondingly. In the phase transition region, both EE and NsEE exhibit dramatic changes. In the paramagnetic phase, as $h$ increases, both EE and NsEE decrease, showing consistent behavior. Additionally, it is evident that EE and NsEE share the same system size dependence. As the system size increases from $4\times 10$ to $6 \times 12$, the phase transition point shifts in the same manner for both EE and NsEE. These results demonstrate that NsEE is also a valuable metric for identifying phase transitions.

We also observe that the NsEE is significantly lower than the EE across both phases.
We can also plot the reduction rate of EE, defined as (\text{EE} - \text{NsEE})/\text{EE} as shown in the inset of Fig.~\ref{Ising}, where the results are for the $6 \times 10$ system. At the critical point, the reduction rate exhibits a peak. Different phases exhibit distinct reduction rates, with the paramagnetic phase showing a higher reduction rate than the ferromagnetic phase. This indicates that the paramagnetic phase possesses more stabilizer-associated entanglement compared to the ferromagnetic phase. Both NsEE and EE reveal that the paramagnetic phase is more challenging to simulate than the ferromagnetic phase, aligning with our expectations.

\subsection{2D XXZ model}
In addition to the 2D transverse Ising model, the 2D XXZ model is another important model in quantum magnetism, particularly useful for studying anisotropic interactions in spin systems. The Hamiltonian for the 2D XXZ model is typically written as:
\begin{equation}
    H_{\text{XXZ}} = J\sum_{\langle i,j \rangle} (S^x_i S^x_j + S^y_i S^y_j + \Delta S^z_i S^z_j)
    \label{H_XXZ}
\end{equation}
where $S^x$, $S^y$ and $S^z$ are the spin-$1/2$ operator for the $x$, $y$ and $z$ direction at site $i$. $\Delta$ is the anisotropy parameter, and the coupling strength $J$ is set to 1 as the unit of energy. $\langle i,j \rangle$ denotes summation over nearest-neighbor pairs.

\begin{figure}[t]
    \includegraphics[width=85mm]{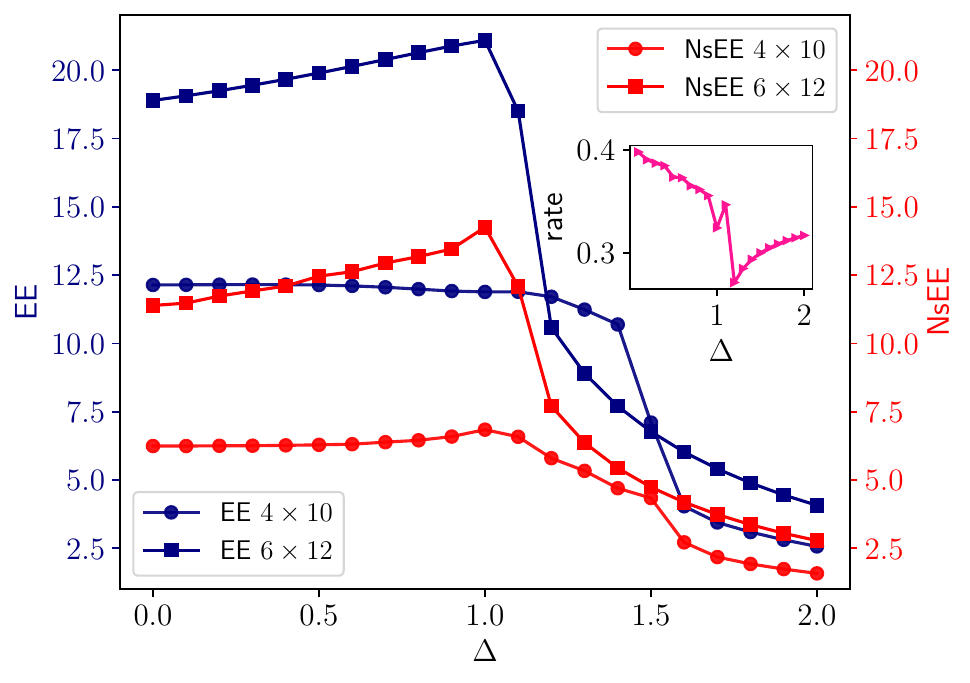}
       \caption{Numerical results for 2D XXZ model with OBC. The entanglement entropy (EE) and the corresponding NsEE of the 2D cuts (Eq.~(\ref{NsEE})) as a function of the $\Delta$ are shown. The insert shows the reduction rate between EE and NsEE for system size $6\times 12$.}
       \label{6XXZ}
\end{figure}

There are two phases in the 2D XXZ model for $\Delta > 0$, i.e., the XY phase for $ 0\leq\Delta < 1.0 $ and the antiferromagnetic phase for $ \Delta > 1.0 $. The critical point is located at $ \Delta = 1.0 $. Similar to the 2D transverse Ising model, one can also use the entanglement entropy to detect this critical point. The numerical results for the 2D XXZ model with system size $4\times10$ and $6\times 12$ under OBC are shown in Fig.~\ref{6XXZ}.

We find that the NsEE is more sensitive to phase transitions than EE in the 2D XXZ model. For instance, the EE for system size of $4\times10$ shows a significant decrease at $\Delta=1.4$, which is far away from the critical point. In contrast, the NsEE begins to decrease at $\Delta=1.0$, which is the actual critical point. Although this change is not very sharp, it still provides a more reasonable and accurate critical point than EE for this small system size. As the system size increases to $6\times12$, the behavior of EE and NsEE becomes consistent. Both EE and NsEE exhibit a sharp decrease at the critical point $\Delta=1.0$. 

The rate of reduction for the $6\times 12$ system of EE is also shown in the inset of Fig.~\ref{6XXZ}. Similar to the 2D transverse Ising model, the reduction rate exhibits a sudden change at the critical point. The value of the reduction rate varies across different phases. The reduction rate is larger in the XY phase compared to the antiferromagnetic phase.

\subsection{Random Clifford + T Circuits}

\begin{figure*}[t]
    \includegraphics[width=55mm]{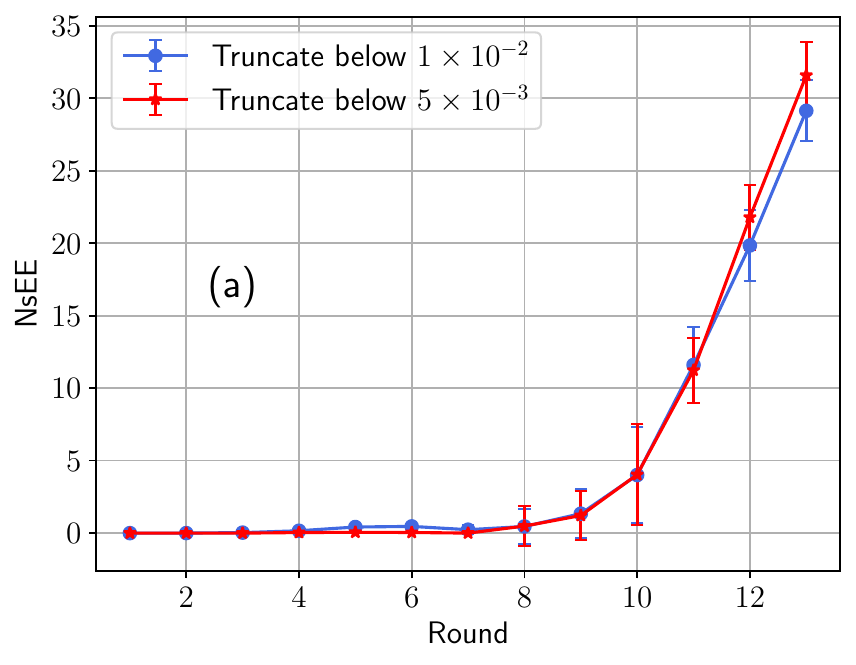}
    \includegraphics[width=55mm]{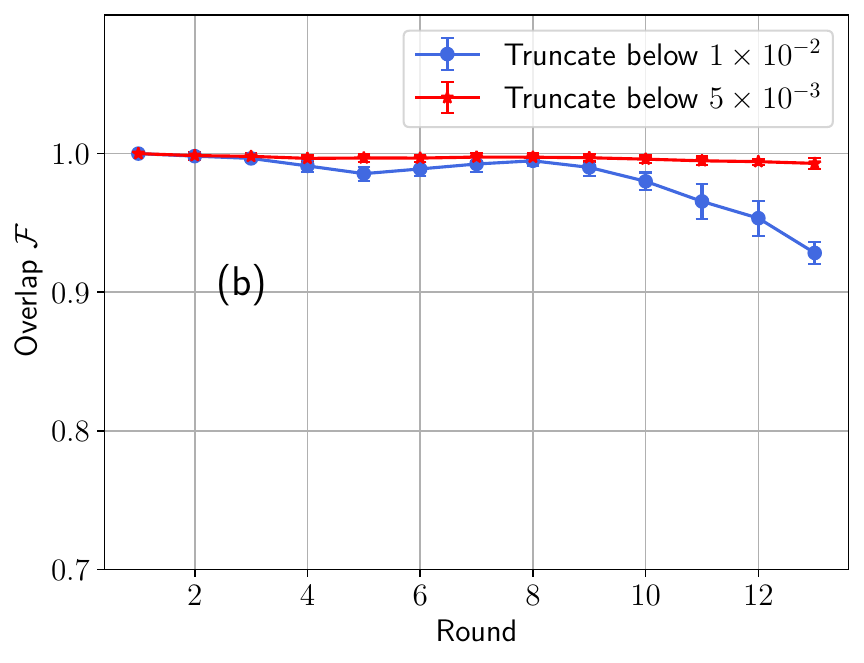}
    \includegraphics[width=55mm]{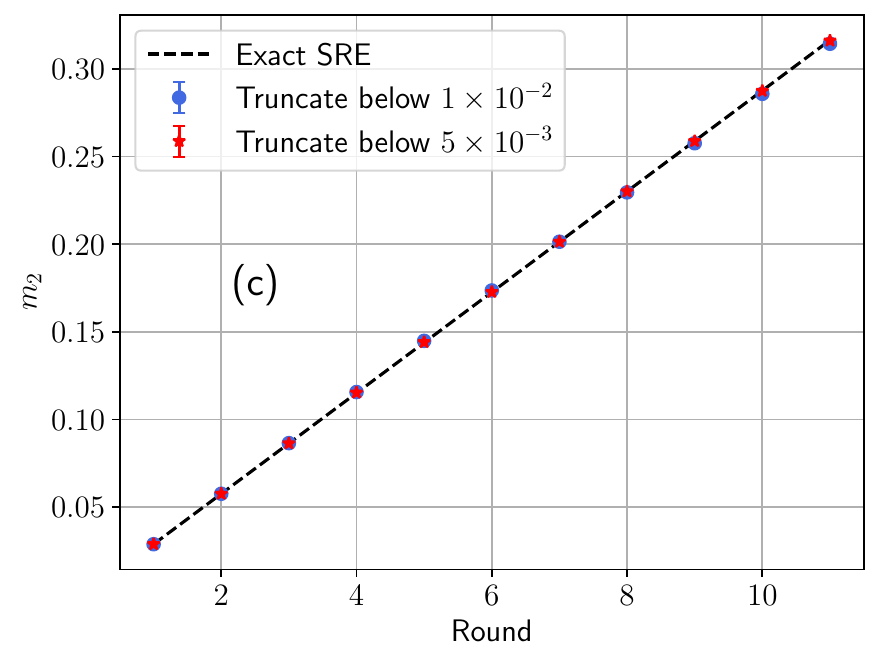}
       \caption{
       NsEE and SRE for random Clifford + $T$ circuits. (a) NsEE against the round for different truncation thresholds. 
       (b) Overlap $\mathcal{F} = \langle \phi| \mathcal{C}|\psi\rangle$ against the round for different truncation thresholds.
       (c) 2nd-order SRE density $m_2$ against the round for truncation thresholds, showing good agreement with the exact values.
       We notice the existence of a transition point at round equals $9$ for both NsEE and overlap. When the round is less than $9$, the NsEE is almost zero, meaning these states can be transformed to nearly product states with Clifford circuits.} 
       \label{Random}
\end{figure*}

\begin{figure}[t]
    \includegraphics[width=80mm]{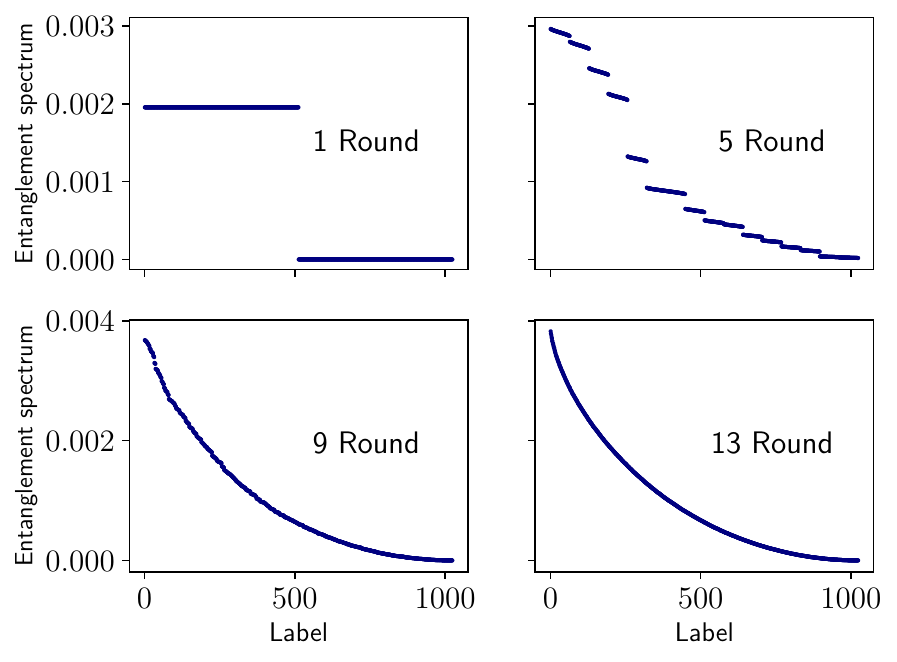}
       \caption{
       Entanglement spectrum of a $N = 20$ qubits system. Starting from the product state $|0\rangle^{\otimes N}$, every round contains 20 layers 2-qubit Clifford circuits to ensure the state is highly entangled, followed by two $T$ gates on two random qubits. The results show that with the increase of $T$ gates, the entanglement spectrum changes from flat to normal.}
       \label{EE_SP}
\end{figure}

We finally consider the simulation of random Clifford + T circuits ~\cite{True2022transitionsin}. In this case, the state is prepared by applying a series of random Clifford circuit $U^{c_i}$ and single-qubit $T$ gates to an initial state $|0\rangle^{\otimes N}$. We consider a system of $N=20$ qubits. Before applying the $T$ gates, we apply $20$ layers of 2-qubit Clifford circuits randomly chosen from the $11520$ possibilities~\cite{10.1063/1.4903507,PhysRevB.100.134306,PhysRevA.87.030301}, to make sure the state is highly entangled. And then we apply two $T$ gates to two different random qubits. Hence the state we prepare is: 
\begin{equation}
    |\psi\rangle= \prod_{i=1}^{R_t} (T^{a_i}  T^{b_i} U^{c_i})|0\rangle^{\otimes N}
    \label{Random_state}
\end{equation}
where $R_t$ is the number of the rounds the Clifford + T circuits are applied, $a_i$ and $b_i$ are the qubits to apply the $T$ gates, and $U^c$ is the Clifford circuit.

Since this state is a random state without a corresponding Hamiltonian, we modify the CA-MPS method to calculate the NsEE. In this case, we directly apply the two-qubit Clifford circuits to the MPS wave function aiming to reduce the entanglement. The criterion for choosing the optimal Clifford circuits is different from the one in CA-MPS. Here we determine the optimal two-qubit Clifford circuits as the one makes the entanglement entropy smallest and then discard the singular values below the truncation threshold. And we perform sweeps same as in DMRG until the calculated NsEE (Eq.~(\ref{NsEE})) is converged.

In practical calculations, we usually obtain the approximated ground state. To mimic this effect, we introduce an SVD truncation process to approximate the exact state in Eq.~(\ref{Random_state}) which is transformed into a MPS representation. 
The SVD truncation is implemented by discarding singular values below a certain threshold, which is set to $5\times 10^{-3}$ and $1\times 10^{-2}$ in our calculations for different runs. We then apply the modified CA-MPS method mentioned above to calculate the NsEE. During this process, we also discard singular values that fall below the set threshold after the SVD decomposition step. To quantify the deviation of the approximated state from the exact state, we calculate the overlap between the two states, defined as:
\begin{equation}
    \mathcal{F} = |\langle \phi|\mathcal{C}|\psi\rangle|
    \label{overlap}
\end{equation}
where $\mathcal{C}$ is the chosen Clifford circuits, and $|\phi\rangle$ is the state after the Clifford circuits are applied. 

Other than NsEE, we also calculate the 2nd-order SRE density $m_2 = M_2/N$ of the state $|\phi\rangle$. The average 2nd-order SRE $M_2$ of state $|\psi\rangle$ is exactly given in Ref~\cite{PhysRevLett.128.050402} as
\begin{equation}
    M_2 (R_t) = -\ln[\frac{4+(2^N - 1)(\frac{-4+3(4^N - 2^N)}{4(4^N-1)})^{(2R_t)}}{3+2^N} ]
    \label{SRE_exact}
\end{equation}

The numerical results are shown in Fig.~\ref{Random}. The NsEE and SRE density $m_2$ are calculated after various rounds of the application of Clifford + $T$ circuits and the final results are averaged over 10 independent runs.
From Fig.~\ref{Random} (a), we can find that NsEE is only slightly affected by the truncation threshold, indicating NsEE is insensitive to the approximation in the studied state.

Interestingly, we identify a transition at the $9$th round, where $18$ $T$ gates are applied. 
For rounds fewer than $9$, the NsEE is nearly zero, and the overlap in Eq.~(\ref{overlap}) is almost equal to $1$. This suggests that these states can be transformed into nearly product states using Clifford circuits, making them easily simulatable on a classical computer. However, when the number of rounds exceeds $9$, the NsEE increases to non-zero values, and the overlap decreases with a large truncation threshold. This suggests that the states become increasingly difficult to simulate on a classical computer after the application of more than $9$ rounds of Clifford + $T$ circuits.

However, SRE density $m_2$ has a different behavior. No transition is found in Fig.~\ref{Random} (c). $m_2$ increases linearly with the number of rounds as expected~\cite{haug2024probingquantumcomplexityuniversal} and the values are well consistent with the exact values. 

In this simulation, the EE should be large in every round because we apply 20 layers of 2-qubit Clifford circuits in each round. Meanwhile, as we discussed, the SRE density increases linearly with each round. This means that both SRE and EE fail to capture the transition point. Instead, NsEE effectively captures the transition of the state from being easy to simulate on a classical computer to computationally more challenging. 

We also calculate the entanglement spectrum of the state $|\psi\rangle$ in Eq.~(\ref{Random_state}). The results are shown in Fig.~\ref{EE_SP}. At the first round, the entanglement spectrum is perfectly flat, i.e., all non-zero eigenvalues are the same, as we prove in Sec.~\ref{Stab_state}. As the number of rounds increases, the entanglement spectrum begins to split into multiple plateaus, with eigenvalues within each plateau being nearly identical, as illustrated in the plot for round five in Fig.~\ref{EE_SP}. Subsequently, the number of plateaus increases, and eventually, the entanglement spectrum changes into a smooth curve when the number of $T$ gates approaches $N$. 

We notice that for the state in Eq.~(\ref{Random_state}), the transition of the entanglement spectrum from flat to smoothness and the transition of NsEE from zero to non-zero values occur almost simultaneously. The possible connection between them needs more investigation.  

\section{Conclusion and Perspective}
\label{Con_Per_Sec}
In this work, we introduce a new metric termed Non-stabilizerness Entanglement Entropy to assess the difficulty of simulating quantum many-body systems classically. NsEE, defined as the minimum residual entanglement entropy after the application of Clifford circuits, integrates entanglement entropy and quantum magic. We show NsEE is a better metric than Stabilizer R\'enyi Entropy. We also calculate NsEE in the ground state of several common many-body models including Toric code, 2D transverse Ising, 2D XXZ models, and also random quantum circuits, take advantage of the Clifford circuits Augmented Matrix Product States~\cite{qian2024augmentingdensitymatrixrenormalization} method. The results of the Toric code model demonstrate the effectiveness of CA-MPS method for the calculation of NsEE. The concept of NsEE provides new insight in the understanding of the difficulty in the classical simulation of quantum many-body systems. NsEE can be also served as a metric in the demonstration of quantum advantage in the future. The accurate evaluation of NsEE is intimately connected to the development of efficient and accurate many-body methods. We hope the introduction of NsEE can stimulate the development of more efficient many-body methods.

\begin{acknowledgments}
 We thank useful discussions with Tao Xiang. The calculation in this work is carried out with TensorKit~\cite{foot7}. The computation in this paper were run on the Siyuan-1 cluster supported by the Center for High Performance Computing at Shanghai Jiao Tong University. MQ acknowledges the support from the National Natural Science Foundation of China (Grant No. 12274290), the Innovation Program for Quantum Science and Technology (2021ZD0301902), and the sponsorship from Yangyang Development Fund.
\end{acknowledgments}

\bibliography{main}

\end{document}